\documentclass[a4paper,11pt]{article}
\pdfoutput=1 
\usepackage{jcappub}
\usepackage[T1]{fontenc} 
\usepackage{booktabs}

\title{\boldmath Non-thermal Dark Matter in $U(1)_{B-L}$ Extension of Inert Doublet Model}

\author[a]{Maien Binjonaid,}
\author[b,c]{Ahmed Elsheshtawy,}
\author[c]{Shaaban Khalil}

\affiliation[a]{Department of Physics and Astronomy,
King Saud University, Riyadh, Saudi Arabia.}
\affiliation[b]{Department of Mathematics, Faculty of Science, Ain Shams University, Cairo 11566, Egypt.}
\affiliation[c]{Center for Fundamental Physics, Zewail City of Science and Technology, 6th of October City, Giza 12578, Egypt.}

\emailAdd{maien@ksu.edu.sa}
\emailAdd{ahmedattia@sci.asu.edu.eg}
\emailAdd{skhalil@zewailcity.edu.eg}

\abstract{We propose an extension of the Inert Doublet Model (IDM) that explains both neutrino masses and dark matter (DM) in the intermediate-mass range by incorporating a \( U(1)_{B-L} \) gauge symmetry. This additional symmetry enables the inclusion of right-handed neutrinos, providing a natural mechanism for neutrino mass generation. While the CP-even component of the inert doublet can serve as a DM candidate, its thermal relic abundance is insufficient to match the observed DM density. To address this, we introduce a non-thermal production mechanism, where a heavy scalar associated with the \( U(1)_{B-L} \) symmetry decays into the inert doublet scalar, yielding a viable relic abundance at low reheating temperatures. We also examine both direct and indirect detection prospects for this DM candidate and assess the model against current experimental constraints. }

\begin{document}
\maketitle
\flushbottom

\section{Introduction} \label{intro}
The Standard Model (SM) of particle physics remains a highly successful theory explaining natural phenomena at high energy scales. However, it lacks an explanation for Dark Matter (DM) and neutrino masses. This motivates addressing such issues by extending the SM with additional symmetries and matter content. 

The Inert Doublet Model (IDM) \cite{Deshpande:1977rw, LopezHonorez:2006gr} is an extension of the SM that addresses some of its limitations, notably the lack of a DM candidate. It represents a specific case within the family of Two-Higgs Doublet Models (2HDM) \cite{Branco:2011iw}, where the additional Higgs doublet, $\Phi_2$, remains "inert", as it neither acquires a Vacuum Expectation Value (VEV) nor couples to SM fermions.

The IDM, stabilized by a discrete $Z_2$ symmetry, provides a viable DM candidate. However, DM with a mass exceeding that of the $W$ boson $(M_W)$ encounters a critical challenge: its high mass leads to highly efficient annihilation, particularly into $W^+W^-$ pairs, resulting in a relic density that is too low. This discrepancy places stringent constraints on the model, as the predicted relic density fails to match the observed DM abundance in the universe, posing a significant challenge to accommodating heavy DM particles within the IDM framework.

The IDM has been extensively studied in the literature \cite{Falaki:2023tyd, Abouabid:2023cdz, Ilnicka:2015jba, Belyaev:2016lok, Eiteneuer:2017hoh, Choubey:2017hsq, Basu:2020qoe, Kalinowski:2020rmb, Ghosh:2021noq, Arhrib:2013ela, Kalinowski:2018ylg}. Regarding DM, a well-known result is that to satisfy the relic density constraint (i.e., to fully account for DM), the mass range of the DM particle falls within two regions: a low mass region where \(m_{DM} \in [55, 75]\) GeV, and a high mass region where \(m_{DM} > 500\) GeV, which requires considerable mass fine-tuning to satisfy the relic density constraint. In contrast, the mass range between \(75\) GeV and \(500\) GeV, known as the intermediate mass region, is ruled out by both the efficiency of annihilation processes and direct detection searches. In this region, only points where \(\Omega h^2\) is subdominant can satisfy the relic density constraints. A comprehensive analysis can be found in \cite{Banerjee:2019luv, Banerjee:2021oxc, Banerjee:2021anv, Banerjee:2021xdp, Banerjee:2021hal, Tsai:2019eqi}.

Theoretical constraints on the IDM arise from vacuum stability, perturbativity, and unitarity \cite{Akeroyd:2000wc}. Vacuum stability requires that the scalar potential be bounded from below, which leads to conditions on the quartic couplings. Perturbativity imposes upper limits on the absolute values of these couplings, typically below \(4\pi\). Unitarity constraints are derived from the S-matrix eigenvalues of various scattering processes, further restricting the allowed parameter space. On the other hand, Electroweak precision observables, particularly the \(S\) and \(T\) parameters, provide significant constraints on the IDM. The contributions to the \(S\) and \(T\) parameters depend on the mass splittings between the inert scalars. These constraints favor small mass splittings, especially for higher scalar masses. Collider searches have placed significant constraints on the IDM parameter space. Data from LEP excludes charged scalar masses below approximately 70–90 GeV. The LHC Higgs measurements further constrain the IDM, particularly through the diphoton decay channel \(h \rightarrow \gamma \gamma\), which can be modified by \(H^\pm\) loops. Additionally, searches for invisible Higgs decays to DM limit the possibility of the decay \(h \rightarrow DM DM \) when $m_{\text{DM}} < m_h/2$. A comprehensive phenomenological analysis of constraints on the IDM was provided in \cite{Belyaev:2016lok} and references therein.
 
The IDM faces challenges in predicting zero neutrino masses, a common issue in 2HDMs. To accommodate neutrino masses, \(U(1)\) symmetries are often introduced. For instance, \cite{Campos:2017dgc} augments the 2HDM with an Abelian gauge group, which avoids Flavor-Changing Neutral Currents (FCNC) and allows for neutrino masses via the seesaw mechanism. This paper focuses on Type I 2HDM, which shares similarities with the IDM but does not address DM. Conversely, \cite{Arcadi:2020aot} examines a Type I 2HDM with an Abelian \(B-L\) symmetry, where a Right-Handed (RH) neutrino serves as a DM candidate in a non-standard cosmology. In the context of the IDM, Ref.~\cite{Arroyo-Urena:2019zah} explores a \(U(1)_X\) extension with a complex singlet, contributing to the DM relic density from both the \(Z'\) gauge boson and the neutral scalar, thereby satisfying Planck bounds. However, neutrino masses are not addressed.

Thus, it is essential to extend the IDM to a model that can effectively address both neutrino masses and the intermediate-mass region of DM. One promising approach involves introducing an additional \(U(1)_{B-L}\) symmetry, which facilitates the inclusion of right-handed neutrinos. In this paper, we will highlight the significance of this extension, as it allows for non-thermal production of DM, providing an alternative explanation for its observed abundance in the universe.

The paper is structured as follows. In Section 2, we briefly introduce the \(U(1)_{B-L}\) extension of the IDM, emphasizing that the CP-even scalar of the inert doublet can serve as a stable candidate for DM, and we show that its thermal relic abundance is extremely low. In Section 3, we explore the non-thermal relic abundance in this model, where a heavy scalar field associated with \(U(1)_{B-L}\) decays into the inert doublet scalar field, resulting in a low reheating temperature. We show that, within the framework of non-thermal relic abundance, DM in the IDM can account for the observational limits in the intermediate mass region, which thermal abundance fails to accommodate. Section 4 is dedicated to the direct and indirect detection of this inert scalar DM. Finally, we present our conclusions and prospects in Section 5.


\section{$U(1)_{B-L}$ Extension of the IDM} \label{sec:core1}

In this section, we review the particle content and the spontaneous symmetry breaking mechanisms in the $U(1)_{B-L}$ extension of the IDM. This extension introduces an additional gauge boson associated with the $B-L$ gauge symmetry. To ensure anomaly cancellation of $U(1)_{B-L}$, three right-handed neutrinos, $\nu_R$, each with a $B-L$ charge of -1, must be included. This setup provides a framework for generating neutrino masses via the seesaw mechanism, where left-handed neutrinos mix with the heavy Majorana neutrinos.

In this class of models, the Higgs sector consists of three scalar fields: an active Higgs doublet, \( \Phi_1 \), which is even under a \( Z_2 \) symmetry, carries \( Y = 1 \) under \( U(1)_Y \), and has \( B-L = 0 \). This doublet is responsible for breaking the electroweak symmetry. There is also an inert Higgs doublet, \( \Phi_2 \), which is odd under the \( Z_2 \) symmetry and has \( Y = 1 \) under \( U(1)_Y \) and \( B-L = 0 \). This doublet does not acquire a vacuum expectation value (VEV) and does not couple to fermions, hence the term "inert." Additionally, a singlet scalar field, \( \eta \), with \( Y = 0 \) and \( B-L = -2 \), triggers the spontaneous breaking of the \( U(1)_{B-L} \) symmetry. The quantum numbers of the particle spectrum in our model are provided in Table \ref{tab:content}.

\begin{table}[ht]
\begin{center}
	\begin{tabular}{ ||c | c c c c c c | c c c|| }
		\hline 
		\hspace{0.5cm} &\hspace{0.05cm} $Q$ \hspace{0.05cm}& \hspace{0.05cm}$u_R$\hspace{0.05cm} &\hspace{0.05cm} $d_R$ \hspace{0.05cm}&\hspace{0.05cm} $L$\hspace{0.05cm} & \hspace{0.05cm}$e_R$ \hspace{0.05cm}& \hspace{0.05cm}$\nu_R $\hspace{0.05cm} &\hspace{0.05cm}$\Phi_1$\hspace{0.05cm} &\hspace{0.05cm}  $\Phi_2$  \hspace{0.05cm} &\hspace{0.05cm}  $\eta$  \hspace{0.05cm} \\ 
		\hline \hline
		$\rm SU(3)_C$ & $3$ & $3$ & $3$ & $1$ & $1$ & $1$ & $1$ & $1$ & $1$ \\
		$\rm SU(2)_L$ & $2$ & $1$ & $1$ & $2$ & $1$ & $1$ &  $2$ & $2$  & $1$\\
		$\rm U(1)_Y$ & $\frac{1}{6}$ & $\frac{2}{3}$ & $-\frac{1}{3}$ & $-\frac{1}{2}$ & $-1$ & $0$ &  $\frac{1}{2}$ & $\frac{1}{2}$& $0$ \\[1mm]
		\hline
		$\rm U(1)_{B-L}$ & $1/3$ & $1/3$ & $1/3$ & $-1$ & $-1$ & $-1$ & $0$ & $0$ & $-2$ \\
		$\rm Z_2$ & $1$ & $1$ & $1$ & $1$ & $1$ & $1$ & $1$ & $-1$ & $1$ \\
		\hline 
		\hline
	\end{tabular}
\end{center}	
	\caption{Particle content and transformation properties under the SM  and  $U(1)_{B-L}$ gauged symmetry.}
	\label{tab:content}
\end{table} 

The Yukawa interactions in this model describe the couplings between the active Higgs doublet and the fermions. As mentioned, the inert doublet does not couple to fermions, while the singlet scalar couples to the right-handed neutrinos. The complete Yukawa Lagrangian is given by:
\begin{equation}
\mathcal{L}_{\text{Yukawa}} = - Y_u^{ij} \overline{Q}_L^i \tilde{\Phi}_1 u_R^j - Y_d^{ij} \overline{Q}_L^i \Phi_1 d_R^j - Y_e^{ij} \overline{L}_L^i \Phi_1 e_R^j - Y_\nu^{ij} \overline{L}_L^i \tilde{\Phi}_1 \nu_R^j - Y_\nu^{ij} \overline{\nu}_R^i \eta (\nu_R^j)^c + \text{h.c.}
\end{equation}
The Higgs potential for the model can be expressed as:
\begin{eqnarray}
V(\Phi_1, \Phi_2, \eta)  &=& -\mu_1^2 |\Phi_1|^2 + \mu_2^2 |\Phi_2|^2 - \mu_\eta^2 |\eta|^2 + \lambda_1 |\Phi_1|^4 + \lambda_2 |\Phi_2|^4 + \lambda_3 |\Phi_1|^2 |\Phi_2|^2  +  \lambda_4 |\Phi_1^\dagger \Phi_2|^2  \nonumber\\ 
  &+& \lambda_\eta |\eta|^4 + \kappa_1  |\eta|^2 |\Phi_1|^2 + \kappa_2 |\eta|^2 |\Phi_2|^2  +  \dfrac{\lambda_5}{2} (\Phi_1^\dagger \Phi_2 \Phi_1^\dagger \Phi_2 +\Phi_2^\dagger \Phi_1 \Phi_2^\dagger \Phi_1)
\end{eqnarray}
It is noteworthy that since \( \Phi_2 \) does not acquire a non-zero VEV, its mass term in the potential should have a positive sign to ensure vacuum stability.
In this regard,  the VEV of $\Phi_1$, $\Phi_2$ and $\eta$ are given by  
\begin{equation}
\langle \Phi_1 \rangle = \begin{pmatrix} 0 \\ v \end{pmatrix}, \quad  \langle \Phi_2 \rangle = \begin{pmatrix} 0 \\ 0 \end{pmatrix}, \quad \langle \eta \rangle = v_\eta.
\end{equation}
As the scale of $U(1)_{B-L}$ symmetry breaking should be much higher than the electroweak scale to account for heavy right-handed neutrino and viable seesaw mechanism, we assume that $v_\eta \gg v$. 
The two Higgs doublets can be parametrized as follows:
 \begin{equation}
 \Phi_1 = \begin{pmatrix}
   G^+ \\
   \frac{1}{\sqrt{2}} (v + h_1 + i G^0)
   \end{pmatrix}, \quad
   \Phi_2 = \begin{pmatrix}
   H^+ \\
   \frac{1}{\sqrt{2}} (\chi + i A)
   \end{pmatrix}
\end{equation}
  where  \(h_1\) and \(\chi\) are the CP-even neutral components, \(G^+\) and \(G^0\) are the Goldstone bosons, \(H^+\) is the charged scalar from the inert doublet, and \(A\) is the CP-odd neutral scalar.  
In addition, the Higgs singlet $\eta$ can be represented as 
 \begin{equation}
   \eta = \frac{1}{\sqrt{2}} (v_\eta + \sigma + i \zeta)
\end{equation}
where \(\sigma\) is the CP-even neutral component, and \(\zeta\) is the CP-odd neutral component.

After symmetry breaking and mixing, the mass matrix for the CP-even neutral scalars in the basis $\{h_1, \sigma, \chi \}$ can be derived from the scalar potential, expressed as:
\begin{equation}
\mathcal{M}^2 = \begin{pmatrix}
2 \lambda_1 v^2 &~~~  \kappa_1 v v_{\eta} & ~~0 \\
\kappa_1 v v_{\eta} &~~~ 2 \lambda_\eta v_\eta^2 & ~~0\\
0 &  0 & \frac{\kappa_2 v_{\eta}^2}{2} + \lambda_l v^2  + \mu_2^2
\end{pmatrix},
\end{equation}
where $\lambda_l \equiv \frac{1}{2}(\lambda_3 + \lambda_4 + \lambda_5)$. Note that there is no mixing between  $h_1$ and $\chi$ with $\sigma$, because $\Phi_2$  does not acquire any VEV. The eigenvalues (mass squared) are:
\begin{equation}
   m_{h,H}^2 = \lambda_1 v^2 + \lambda_\eta v_{\eta}^2 \mp \sqrt{(\lambda_1 v^2 - \lambda_\eta v_{\eta}^2)^2 + (\kappa_1 v v_{\eta})^2}
\end{equation}
 The mass eigenstates are mixtures of \(h\) and \(\chi\) with a mixing angle \(\theta\):
\begin{equation}
   \tan 2\theta = \frac{\kappa_1 v v_{\eta}}{\lambda_1 v^2 - \lambda_\eta v_{\eta}^2}
\end{equation}
While the mass of $\chi$, which is stable and acts as a candidate for DM, and the mass of pseudoscaler $A$ are given by 
\begin{align}
    m_{\chi}^2 &= \mu_2^2 + \lambda_l v^2 + \frac{\kappa_2 v_{\eta}^2}{2}\\
m_{A}^2 &=\mu_2^2 + \overline{\lambda_l} v^2 + \frac{\kappa_2 v_{\eta}^2}{2},
\end{align}
 with $\overline{ \lambda_l} \equiv \frac{1}{2}(\lambda_3 + \lambda_4 - \lambda_5)$.
The lightest mass eigenstate, \(h\), is identified as the SM-like Higgs boson, with a mass of \(m_{h} = 125\) GeV. In contrast, \(H\) represents the heaviest CP-even neutral scalar, with a mass approximately equal to \(m_{H} \sim v_\eta\). The interaction coupling \(\lambda_{H \chi \chi}\) between the heavy CP-even neutral mass eigenstate \(H\) and two copies of the inert CP-even scalar \(\chi\), which describes the strength of the decay of the heavy \(H\) to two DM particles, is given by:
\begin{equation}
\lambda_{H \chi \chi} = \lambda_3 v \sin^2 \theta + \kappa_2 v_\eta \cos^2 \theta
\end{equation}

Assuming the standard thermal evolution of the Universe, the relic abundance \(\Omega h^2\) of inert DM is inversely proportional to the annihilation cross-section, as described by the following equation:
\begin{equation}
\Omega h^2 = \frac{1.07 \times 10^9 \text{ GeV}^{-1}}{M_{\text{Pl}} \sqrt{g_*} \langle \sigma v \rangle }, 
\end{equation}
where \( M_{\text{Pl}} \) is the Planck mass (\( M_{\text{Pl}} \approx 1.22 \times 10^{19} \text{ GeV} \)), \( g_* \) is the effective number of relativistic degrees of freedom at the freeze-out temperature, \( \langle \sigma v \rangle \) is the thermal average of the annihilation cross-section multiplied by the relative velocity.

For \(\chi\) with a mass greater than the \(W\) boson mass, the annihilation cross section is predominantly driven by the process \(\chi \chi \to W^+ W^-\). This process significantly enhances the annihilation rate, leading to a large cross section and, consequently, a very low relic density of the IDM. Additionally, a region of efficient annihilation is observed near \(m_h/2\), where the relic density drops sharply due to resonant enhancement of the annihilation cross section. On the other hand, when \(\chi\) has a mass below the \(W\) boson mass, the annihilation cross section becomes less efficient. This inefficiency results in a significantly higher relic abundance of the IDM, often exceeding the observational upper limits by a considerable margin.  

\begin{figure}[h!]
\centering
\includegraphics[scale=0.34]{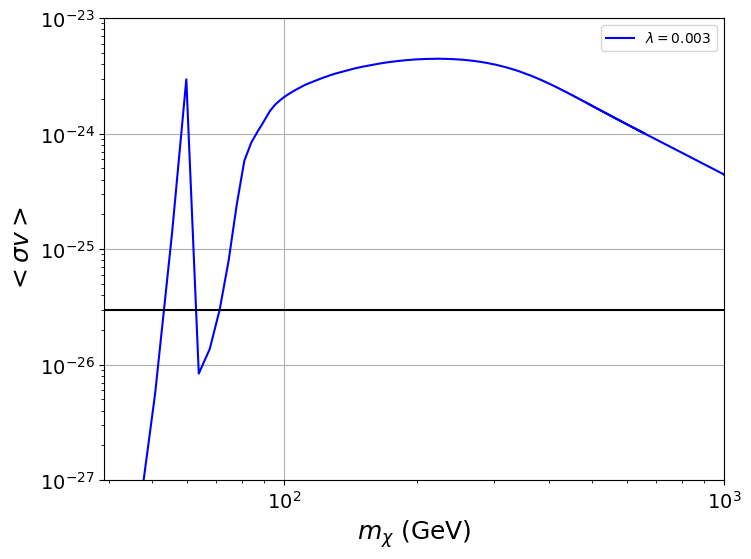}
\includegraphics[scale=0.34]{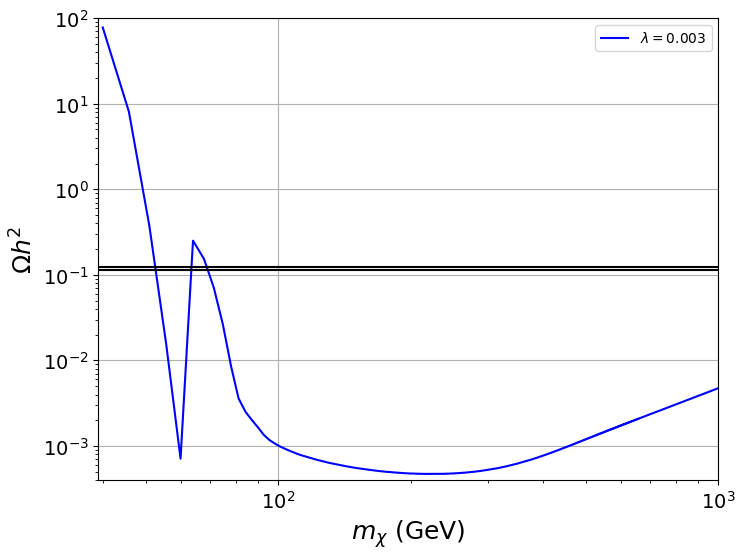}
\caption{The thermally-averaged annihilation cross-section (in cm\(^3\)/s), left panel, and the thermal relic abundance \(\Omega h^2\), right panel, as functions of the DM mass in the IDM.}
\label{without_TR}
\end{figure}

These conclusions are illustrated in Fig. \ref{without_TR}, which displays the annihilation cross section and the thermal relic abundance of the IDM, \(\Omega_\chi h^2\), as functions of its mass. The plots were generated using \texttt{micrOMEGAs} v.6 \cite{Alguero:2023zol}.

\section{Non-thermal Relic Abundance in IDM}

In the context of cosmology, particularly in scenarios involving the breaking of the \(U(1)_{B-L}\) symmetry and the decay of a heavy scalar field \(H_2\), the reheat temperature \(T_{\text{RH}}\) is an important quantity that describes the temperature of the universe once it thermalizes following such a decay. The reheating temperature can be related to the decay width \(\Gamma_{H_2}\) of the scalar field \(H_2\). The general formula for the reheat temperature \(T_{\text{RH}}\) is given by:
\begin{equation}
T_{\text{RH}} \approx \left(\frac{90}{\pi^2 g_*}\right)^{1/4} \sqrt{\Gamma_{H_2} M_{\text{Pl}}},
\label{trh}
\end{equation}
where $T_{\text{RH}}$ is the reheat temperature, $g_*$ is the effective number of relativistic degrees of freedom at the time of reheating, and $M_{\text{Pl}}$ is the reduced Planck mass.

In this case, the abundance of DM candidate \( \chi \) can be determined by solving a system of Boltzmann equations that describe the evolution of the scalar field \( H_2 \), the DM \( \chi \), and radiation \( R \). 
For simplicity and to align with standard conventions, we will refer to the heavy scalar \( H_2 \) as \( \phi \).
The coupled Boltzmann equations, assuming kinetic equilibrium, are:
\begin{eqnarray}
\frac{d\rho_\phi}{dt}&=&- 3H \rho_\phi-\Gamma_\phi \rho_\phi \\
\frac{d\rho_r}{dt}&=&- 4H \rho_r+\Gamma_\phi \rho_r(1-\frac{b \langle E_\chi \rangle}{m_\phi}) +  \langle \sigma v \rangle  2  \langle E_\chi \rangle [ n_\chi^2- (n_\chi^{eq})^2] 
\label{rhor} \\
\frac{dn_\chi}{dt}&=&- 3H n_\chi+\frac{b}{m_\phi} \Gamma_\phi \rho_\phi-  \langle \sigma v \rangle    [ n_\chi^2- (n_\chi^{eq})^2]  
\label{dm}
\end{eqnarray}
Where $\rho_\phi, \rho_r$ are the energy density of the scalar field and radiation, respectively, and $n_\chi$ is the number density of the DM.  $\langle E_\chi \rangle \simeq \sqrt{m_\chi^2+3 T^2}$ is the average energy of the DM. The second term of the right side of eq. (\ref{dm}) denotes the DM non-thermal production from the scalar decay, while the third term is the DM annihilation. Here, $b$ is the branching ratio of the scalar decaying into the DM sector. 
To solve the Boltzmann equation, we define the following dimensionless parameters:
\begin{eqnarray}
\Phi \equiv \rho_\phi T_{\rm RH}^{-1} a^3;  R \equiv \rho_r a^4 ;  X \equiv n_\chi a^3;  A \equiv a T_{\rm RH}
\end{eqnarray}
The Boltzmann equation can be rewritten as \cite{Drees:2017iod,Han:2019vxi}
\begin{eqnarray}
 \tilde{H}\frac{d\Phi}{dA}&=& - c_\rho^{1/2} A^{1/2} \Phi  \\
 \tilde{H}\frac{dR}{dA}&=& c_\rho^{1/2} A^{3/2} (1-\frac{b \langle E_\chi \rangle}{m_\phi}) \Phi   
+c^{1/2}_1 A^{-3/2}  \langle \sigma v \rangle  2\langle E_\chi \rangle M_{Pl} (X^2-X_{eq}^2)~~~~~~~~~   \label{R} \\
 \tilde{H}\frac{dX}{dA}&=&  A^{1/2} T_{\rm RH} \Phi \frac{b}{m_\phi}   
  -c^{1/2}_1 A^{-5/2} \langle \sigma v \rangle M_{Pl} T_{\rm RH} (X^2-X_{eq}^2)
 \end{eqnarray}
where $c_\rho= \frac{\pi^2 g_*(T_{\rm RH})}{30}$, $c_1= \frac{3}{8\pi}$. Finally,
the scaled $X$ equilibrium number density $X_{\rm eq}$ is given by
\begin{equation} \label{eq:X_eq}
X_{\rm eq} \equiv \left( \frac{A} {T_{\rm RH}} \right)^3 
\frac{g_{X} T {M_{X}}^2} {2 \pi^2} K_2 \left( 
\frac{M_{X}} {T} \right)  \rightarrow 
{ 
\left\{ \begin{array}{l}
\left( \frac{A} {T_{\rm RH}} \right)^3 \frac{ g_X \,\zeta(3) T^3} {\pi^2}
\hspace*{2.9cm} {\rm if} \ T \gg M_{X} \\
\left( \frac{A} {T_{\rm RH}} \right)^3 g_{X} 
\left( \frac {M_{X} T} {2 \pi} \right)^{\frac{3}{2}} \exp(-M_{X}/T) \ \ 
{\rm if} \ T \ll M_{X} 
\end{array} 
\right.}
\end{equation}
Here $g_{X}$ counts the internal degrees of freedom of $X$, and $K_2$ is the modified Bessel function of the second kind.

Given that the energy density of the universe was primarily dominated by the scalar field $\phi$ before its decay began, it is naturally to assume the following initial conditions.
\begin{equation} 
\Phi_I = \frac{3 M^2_P H^2_I}{T^4_{RH}}, R_I =
X_I = 0, A_I =1.
\end{equation}

In the  period between $H^{-1}_I$ and $\Gamma^{-1}_{\phi}$ (indicating the completion of the $\phi$ decay)
the Hubble expansion parameter $H$ can be written as~\cite{Giudice:2000ex}
\begin{equation} 
H = \sqrt{\frac{5 \pi^2
g_*^2(T)} {72 g_*(T_{RH})}} \frac{ T^4} {T_{\rm
RH}^2 M_{P}}, %
\label{H-defn}
\end{equation}
and the temperature is inferred from the relation %
\begin{equation} %
T = \left(\frac{30}{ \pi^2 g_*(T)} \right)^{1/4} \frac{R^{1/4}}{A}
T_{RH}.%
\end{equation}%

\begin{figure}[h!]
\centering
\includegraphics[width=9cm,height=5cm]{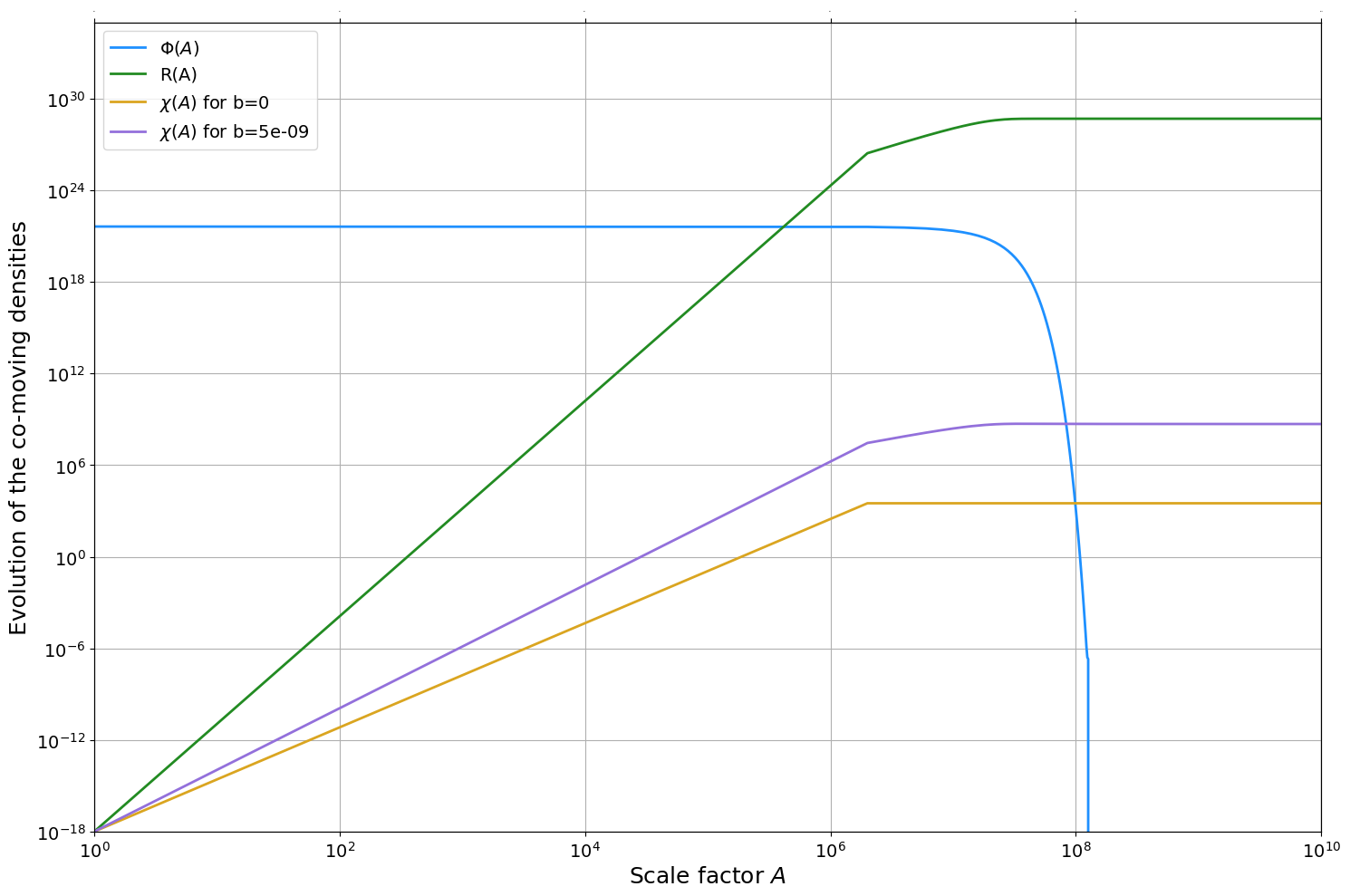}
\caption{Evolution of the co-moving densities $\Phi$, $R$ and $\chi$, as well as $\chi_{eq}$ as function of the normalized scale $A$, for $T_{RH} = 1 \rm GeV$, $m_{\chi}=200 \rm GeV, $ $b=0, 5 \times 10^{-9}$.}
\label{evolution}
\end{figure}

Figure \ref{evolution} illustrates the evolution of the co-moving densities $X$ and  $\Phi$, as derived from the solutions to the above Boltzmann equations, alongside the equilibrium co-moving density 
$X_{eq}$, all plotted as functions of the normalized scale factor $A$.

The Boltzmann equation for $X$, whose solution will give us the expression for the relic abundance $\Omega_{X} h^2$ of DM in terms of the parameters appearing in the Boltzmann equations. More precisely, the $X$ relic abundance is given by\cite{Drees:2017iod}:
  \begin{equation}\label{omega1}
  \Omega_{DM} h^2 = \frac{\rho_{X}(T_f)}{\rho_{R}(T_f)} \frac{T_f}{T_{\rm now}} \Omega_{R} h^2 = M_{X} \frac{X(T_{f})}{R(T_f)} \frac{A_f T_{f}}{T_{\rm now} T_{RH}} \Omega_{R} h^2 \, .
  \end{equation} 
The present observational
values of the current temperature and density of photons cosmic
microwave background (CMB), as collected by the Particle Data Group, are:
\begin{eqnarray} \label{eq:T_now}
\Omega_{\gamma} h^{2}&=& 2.473\times10^{-5}\,; \\ \nonumber
T_{\rm now}&=&2.7255 \ {\rm K}= 2.35\times 10^{-13}\ {\rm GeV}\,.
\end{eqnarray}
We use these values in our numerical calculations. Cosmological
observations also determine the total present density of non-baryonic
dark matter quite accurately \cite{Planck:2018vyg}:
\begin{equation} \label{eq:Om_now}
\Omega_{DM} h^2 = 0.120 \pm 0.001\,.
\end{equation}

\begin{figure}[h!]
\centering
\includegraphics[width=10cm,height=6cm]{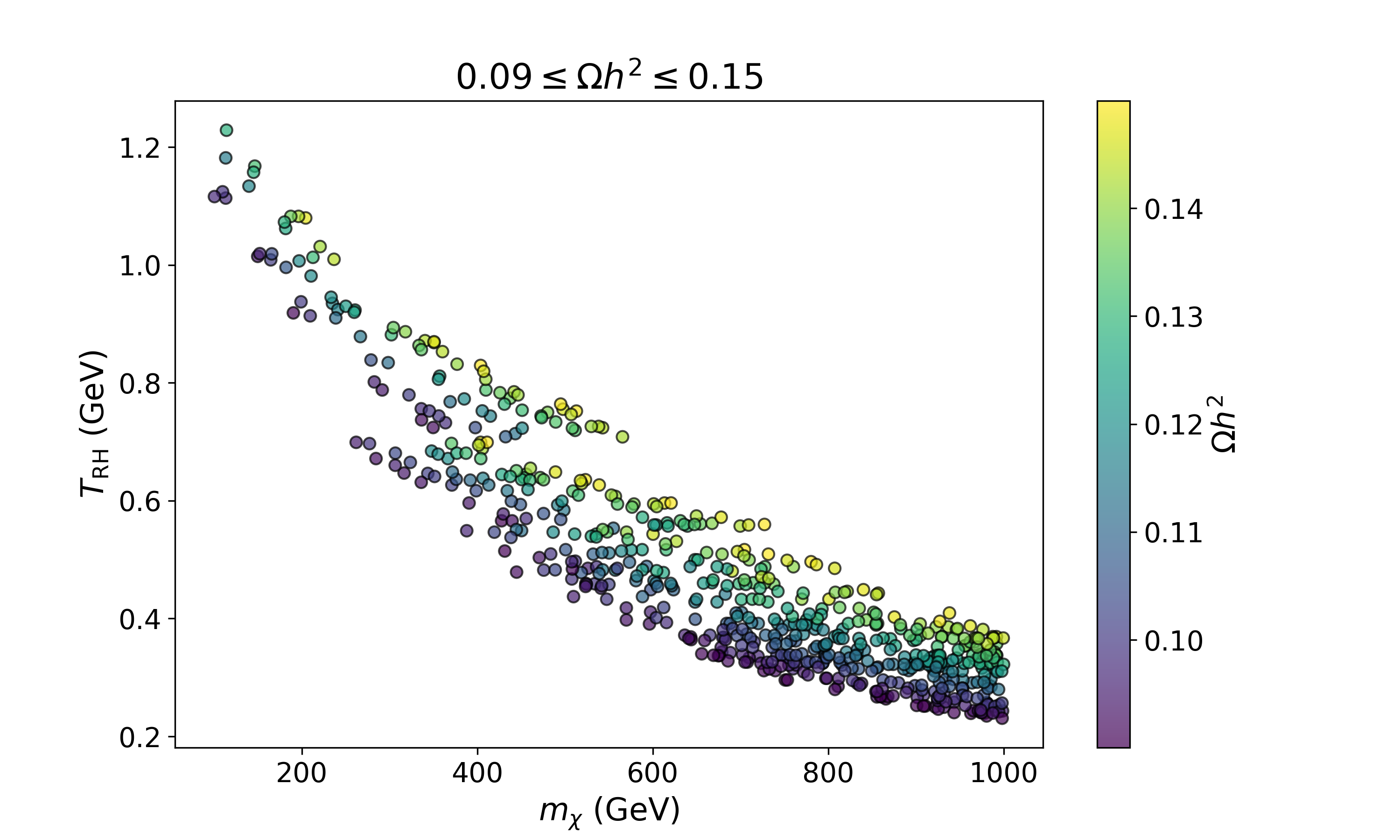}
\caption{Relic abundance of the dark matter ($\chi$) as a function of the reheat temperature for different values of $m_{\chi}$, in the presence of a decaying field $ m_\phi = 10 ~\rm TeV$ with $ b= 1 \times 10^{-9} $.}
\label{omega1}
\end{figure}

The non-thermal production mechanism described above makes it possible to achieve the required relic density in our model as shown in Figure~\ref{omega1}, which presents a scatter plot illustrating how DM mass (on the horizontal axis) and the reheat temperature (on the vertical axis) vary across different parameter points. Each point is colored according to the predicted relic density, as indicated by the color bar on the right.  In particular, for our choices of $m_\phi = 10$ TeV, and setting $b= 1\times 10^{-9}$, we observe that the intermediate region ($100 \leq m_\chi \ \text{(GeV)} \leq 500$) admits values of $\Omega h^2$ within the range of observations. Here, we allow for some deviation from the central value to account for theoretical uncertainties in the computation of $\Omega h^2$. We note that for smaller values of the mass, $T_{\text{RH}}$ that corresponds to the allowed region approaches 1.2 TeV, while dropping by half for $m_\chi \sim 500$ GeV, and continues to drop as the mass increases. On the other hand, we note that for a given value of $m_\chi$, as $T_{\text{RH}}$ increases, $\Omega h^2$ also increases.
It is worth mentioning that if one increases the value of $b$, then this would result in overabundance. While this can be compensated by controlling $m_H$ and $\langle \sigma v \rangle$ (e.g. through controlling the mass splitting), we restricted our analysis to the specified values for practical reasons to find stable numerical solutions.

The points appearing in Figure~\ref{omega1} are selected from a sample of 20K points, where we vary the input parameter $\kappa_2$ between $10^{-5}$ and 0.02. This results in the ranges of the masses listed in Table \ref{tab:params} 

\begin{table}[h!]
\centering
\begin{tabular}{ll}
\toprule
\textbf{Parameter} & \textbf{Range} \\
\midrule
\(m_{\mathrm{DM}}\)     & 40 -- 1000 (\text{GeV}) \\
\(m_{A}\)               & 350 -- 1060 (\text{GeV})\\
\(m_{H^\pm}\)           & 178 -- 1015 (\text{GeV}) \\
\bottomrule
\end{tabular}
\caption{Parameter Ranges}
\label{tab:params}
\end{table}
All other parameters are fixed: $m_h = 125$ GeV, $m_H = 10^4$ GeV, $\lambda_1 = 0.129$, $\lambda_2=0.01$, $\lambda_3 = 1.003$, $\lambda_4 = 1$, $\lambda_5 = -2$, $\lambda_\eta = 0.5$, $\kappa_1 = 0.01$,  $\mu_2^2 = 1000$ GeV$^2$. 
We have checked theoretical constraints via interfacing with \texttt{2HDMC} \cite{Eriksson:2009ws}. The data presented in Figure 3, passed constraints from tree-level unitarity, perturbativity, and vacuum stability. The data also passed constraints from direct detection as computed by \texttt{micrOMEGAs}.  

It is worth noting that while entropy injection from heavy scalar decays generally leads to a reduction in the DM relic density, the direct production of DM from the decay (proportional to the branching ratio \(b\)) can compensate for this effect. For sufficiently large \(b\), the non-thermal production of DM can dominate, resulting in an overall increase in the relic density.

\section{Direct and Indirect Detection of Inert Scalar Dark Matter}

\subsection{Direct Detection}
In this section, we analyze the spin-independent scattering cross-section between DM particles ($\chi$) and nucleons (N) via exchanging SM Higgs bosons. This $\chi$-$N$ scattering process is relevant for the direct detection of DM, which is considered one of the major methods to either detect or constrain DM. The cross-section is given by,
\begin{equation}
 \sigma_{\chi N} = \frac{m_r^2}{4 \pi} \left( \frac{\lambda_l}{m_h^2 m_{\chi}} \right)^2 f^2 m_N^2,
\label{eq:ddtree}
\end{equation}
where $m_N$ is the mass of the nuclei and $f \approx 0.3$ is a nuclear form factor defined in \citep{Elsheshtawy:2023wza, Ellis:2018dmb}.

\begin{figure}[h!]
\centering
\includegraphics[width=10cm,height=7cm]{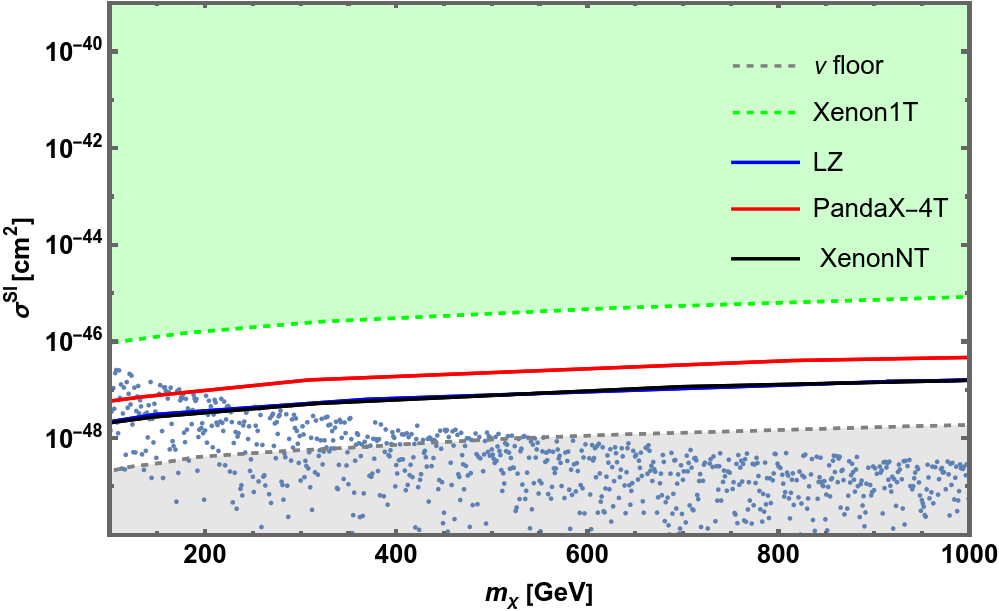}
\caption{The spin-independent cross section of the scattering between the lightest scalar inert DM and nucleon as a function of its mass.}
\label{DDvfloor}
\end{figure}

In Figure~\ref{DDvfloor}, we show the spin-independent scattering cross-section between $\chi$ and nucleon (proton) as a function of $m_{\chi}$ for $m_{h}$ with trilinear coupling $\lambda_l \in [10^{-5},0.003]$. The red, blue, and black curves represent the limits from PandaX-4T \cite{PandaX-4T:2021bab}, LZ \cite{LZ:2024zvo}, and XenonNT \cite{XENON:2024wpa}, respectively. The green shaded area represents the exclusion limit set by the XENON1T experiment \cite{XENON:2017vdw}. The gray shaded area limit-based neutrino floor calculation shown by the dashed line is taken from the recent APPEC report \cite{Billard:2021uyg}. 

As can be seen from this figure, in the region of interest for the DM mass (i.e. $ 100 < m_{\chi} (\text{GeV}) < 500$), the cross-section $\sigma^{SI}$ can be lower than the current DD limits where its value is below $2 \times 10^{-48}$ cm$^2$, while still being above the neutrino floor. As $m_\chi$ increases above 500 GeV, all points in our dataset lie below the neutrino floor, while points below 100 GeV in our dataset are problematic with respect to the theoretical constraints mentioned in the previous Section.

\subsection{Indirect Detection}
One of the main methods to observe the effects of DM is through its annihilation at the core of the galaxy into SM states, which can subsequently lead to an excess of positron flux near Earth. The dominant annihilation channels are the $W^+W^-$ and $ZZ$ bosons for the mass range of interest, and the fragmentation of these bosons into positrons can be fitted by the functions presented in Ref.~\cite{Cirelli:2008id} (defining $x=\frac{E_e^+}{m_{\chi}}$):

\begin{align}
\frac{dN_{e^+}}{d\ln x}\Bigg|_{W^+W^-} &= \exp\Bigg( -1.895 
+ \frac{2.821}{1!} \ln x 
+ \frac{6.299}{2!} (\ln x)^2 
+ \frac{7.563}{3!} (\ln x)^3 \nonumber \\
&\quad + \frac{6.914}{4!} (\ln x)^4 
+ \frac{4.812}{5!} (\ln x)^5 
+ \frac{2.367}{6!} (\ln x)^6 
+ \frac{0.7273}{7!} (\ln x)^7 
+ \frac{0.1050}{8!} (\ln x)^8 \Bigg) 
\end{align}

\begin{align}
\frac{dN_{e^+}}{d\ln x}\Bigg|_{ZZ} &= \exp\Bigg( -2.485 
+ \frac{2.809}{1!} \ln x 
+ \frac{5.501}{2!} (\ln x)^2 
+ \frac{4.901}{3!} (\ln x)^3 \nonumber \\
&\quad + \frac{2.953}{4!} (\ln x)^4 
+ \frac{1.252}{5!} (\ln x)^5 
+ \frac{0.3424}{6!} (\ln x)^6 
+ \frac{0.04574}{7!} (\ln x)^7 \Bigg).
\end{align}

The positron flux near Earth is given by,
\begin{equation}
 \Phi_{e^+}(E)
 =
 \frac{c}{4\pi}f_{e^+}(E,\vec{r}_\odot)~,
\end{equation}
where c is the speed of light, and $f_{e^+}(E,\vec{r}_\odot)$ is the number density of positron per energy, which in turn depends on the source term,
\begin{equation}
 Q(E,\vec{r})
 =
 \frac{1}{2} n^2(\vec{r})
 \sum_{f}
 \langle \sigma v \rangle _f
 \left(\frac{d N_{e^+}}{d E}\right)_f~~,
\end{equation}
where, $n$ is the number density of DM, $\langle \sigma v \rangle_f$ is the thermally averaged annihilation cross section to the state $f$. In particular, $f_e^+$ is considered as the solution of the diffusion equation under the steady state condition,

\begin{equation}
 K(E)\nabla^2 f_{e^+}(E,\vec{r})
 +
 \frac{\partial}{\partial E}
 \left[b(E)f_{e^+}(E,\vec{r})\right]
 +
 Q(E,\vec{r})=0~,
 \label{eq:diffusion}
\end{equation}
where $K(E)$ is the diffusion
constant, and $b(E)$ is the energy loss rate. We follow the procedure described in \cite{Hisano:2005ec}, and solve the equation numerically in cylindrical coordinates. 
The background, representing the flux of primary electrons, and secondary electrons and positrons, is given by \cite{Moskalenko:1997gh, Baltz:1998xv},
\begin{eqnarray}
\Phi_{e^-}^{\text{prim}}(E) &=& \frac{0.16 \cdot E^{-1.1}}{1 + 11.0 \cdot E^{0.9} + 3.2 \cdot E^{2.15}} \\
\Phi_{e^-}^{\text{sec}}(E) &=& \frac{0.70 \cdot E^{0.7}}{1 + 110.0 \cdot E^{1.5} + 600.0 \cdot E^{2.9} + 580.0 \cdot E^{4.2}} \\
\Phi_{e^+}^{\text{sec}}(E) &=& \frac{4.5 \cdot E^{0.7}}{1 + 650.0 \cdot E^{2.3} + 1500.0 \cdot E^{4.2}}.
\end{eqnarray} 

\begin{figure}[h!]
    \centering
    \begin{minipage}[b]{0.49\textwidth}
        \centering
        \includegraphics[width=\textwidth]{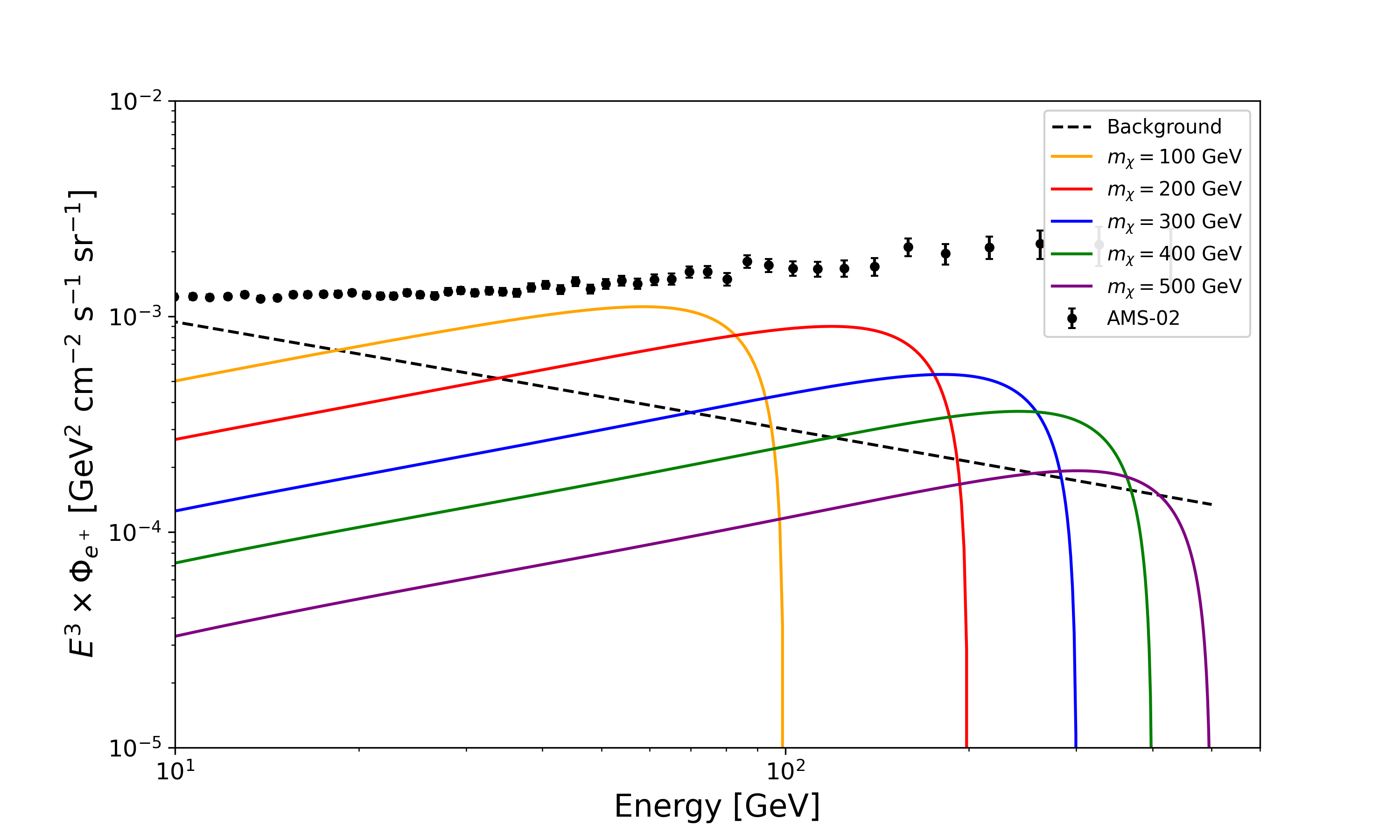}
    \end{minipage}
    \hfill
    \begin{minipage}[b]{0.49\textwidth}
        \centering
        \includegraphics[width=\textwidth]{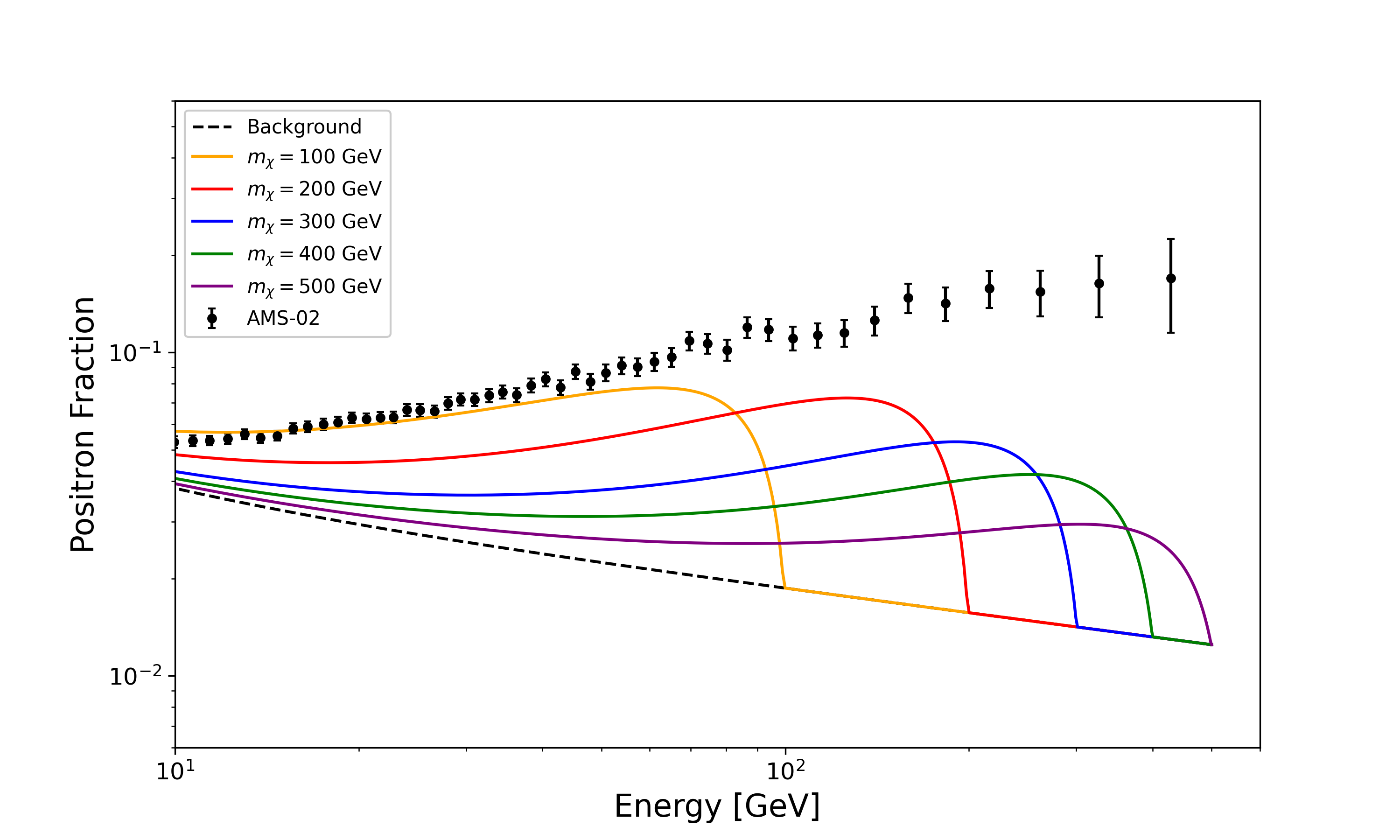}
    \end{minipage}
    \caption{Positron Flux (left) and Positron Fraction (right) as a function of energy for different $m_\chi$ values. Normalization and boost factors are set to 1.}
    \label{ID}
\end{figure}

Figure~\ref{ID} shows the positron flux (left panel) and positron fraction (right panel) as a function of energy for $m_\chi = [100-500]$ GeV. The corresponding contributions to $\langle \sigma v \rangle$, which were computed using micrOMEGAs v.6 \cite{Alguero:2023zol}, are shown in Table~\ref{tabid}.

\begin{table}[!htbp]
\centering
\begin{tabular}{|c|c|c|}
\hline
$m_{\chi}$ [GeV] & $\langle \sigma v \rangle_{WW}$ [$\text{cm}^3/\text{s}$] & $\langle \sigma v \rangle_{ZZ}$ [$\text{cm}^3/\text{s}$] \\
\hline
100 & $1.59 \times 10^{-24}$ & $8.82 \times 10^{-25}$ \\
200 & $1.62 \times 10^{-24}$ & $3.13 \times 10^{-24}$ \\
300 & $1.03 \times 10^{-24}$ & $3.46 \times 10^{-24}$ \\
400 & $6.57 \times 10^{-25}$ & $3.63 \times 10^{-24}$ \\
500 & $4.54 \times 10^{-25}$ & $2.38 \times 10^{-24}$ \\
\hline
\end{tabular}
\caption{$m_{\chi}$ values and their corresponding $\langle \sigma v \rangle$ for $WW$ and $ZZ$ channels.}
\label{tabid}
\end{table}

In Figure~\ref{ID}-(a), the background flux from the positrons is depicted as a dashed line. We observe that for $m_\chi = 100$ GeV, the signal exceeds the background in the range $E=[18-95]$ GeV with $E^3 \Phi_{e^+}$ in the range $[7.6-4.1] \times 10^{-4} \ GeV^2 cm^{-2} s^{-1} sr^{-1}$. As for $m_\chi = 200$ GeV, the signal ranges between $[6-3.2] \times 10^{-4} \ GeV^2 cm^{-2} s^{-1} sr^{-1}$, for energy between 34 to 194 GeV. Next, for $m_\chi = 300$ GeV, the signal flux starts to exceed the background at $E=70$ GeV, reaching a maximum value of $E^3 \Phi \sim 5 \times 10^{-4} \ GeV^2 cm^{-2} s^{-1} sr^{-1}$ at $E \sim 200$ GeV. As for the case of $m_\chi = 400$ GeV, the signal passes through the background at slightly above $E= 100$ GeV, and maximizes to a value of $2.5 \times 10^{-4} \ GeV^2 cm^{-2} s^{-1} sr^{-1}$ at $E=300$ GeV. Lastly, for $m_\chi = 500$ GeV, the signal barely exceeds the backgound between $ 300 \leq E \leq 400$ GeV, with a maximum value of $1 \times 10^{-4} \ GeV^2 cm^{-2} s^{-1} sr^{-1}$.

\begin{figure}[h!]
    \centering
    \begin{minipage}[b]{0.49\textwidth}
        \centering
        \includegraphics[width=\textwidth]{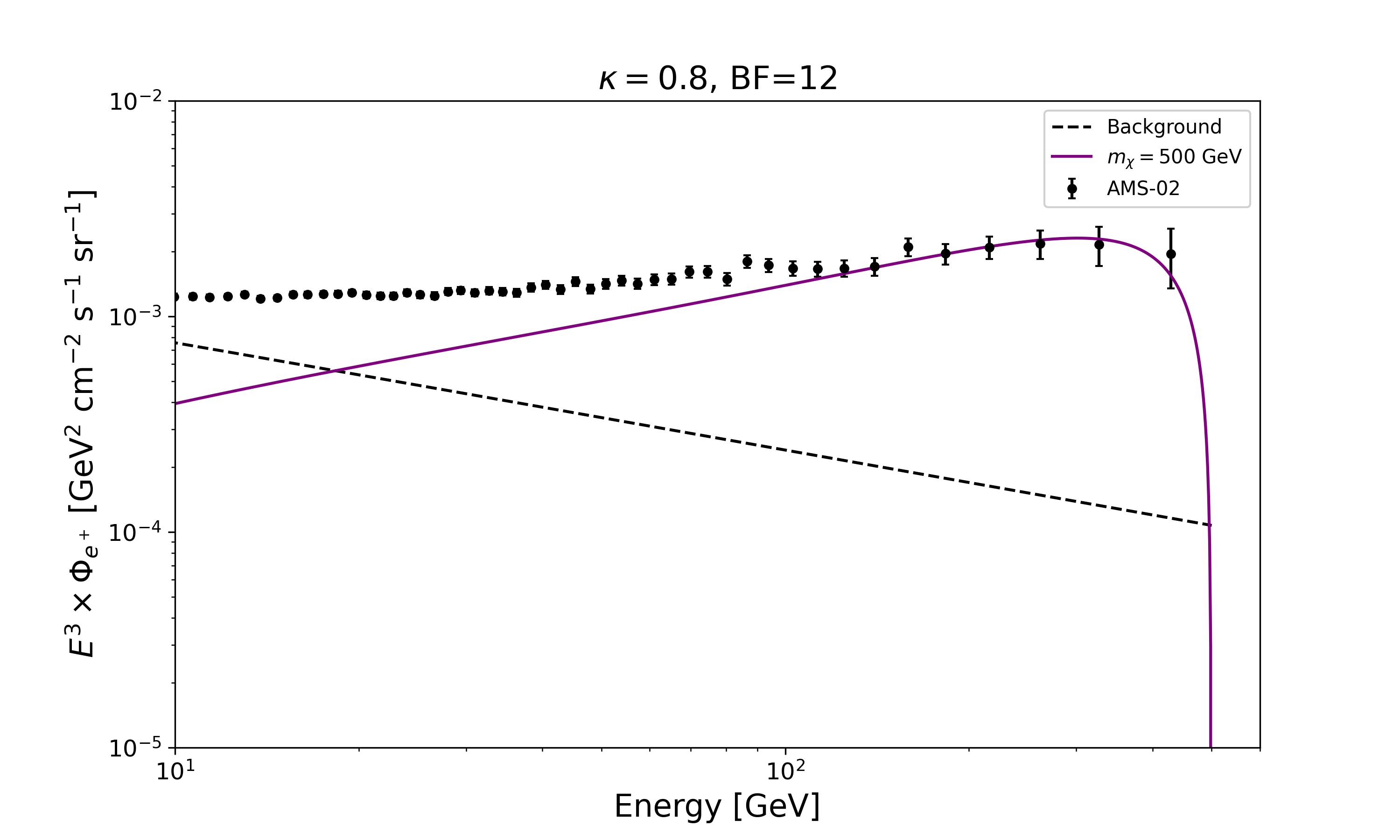}
    \end{minipage}
    \hfill
    \begin{minipage}[b]{0.49\textwidth}
        \centering
        \includegraphics[width=\textwidth]{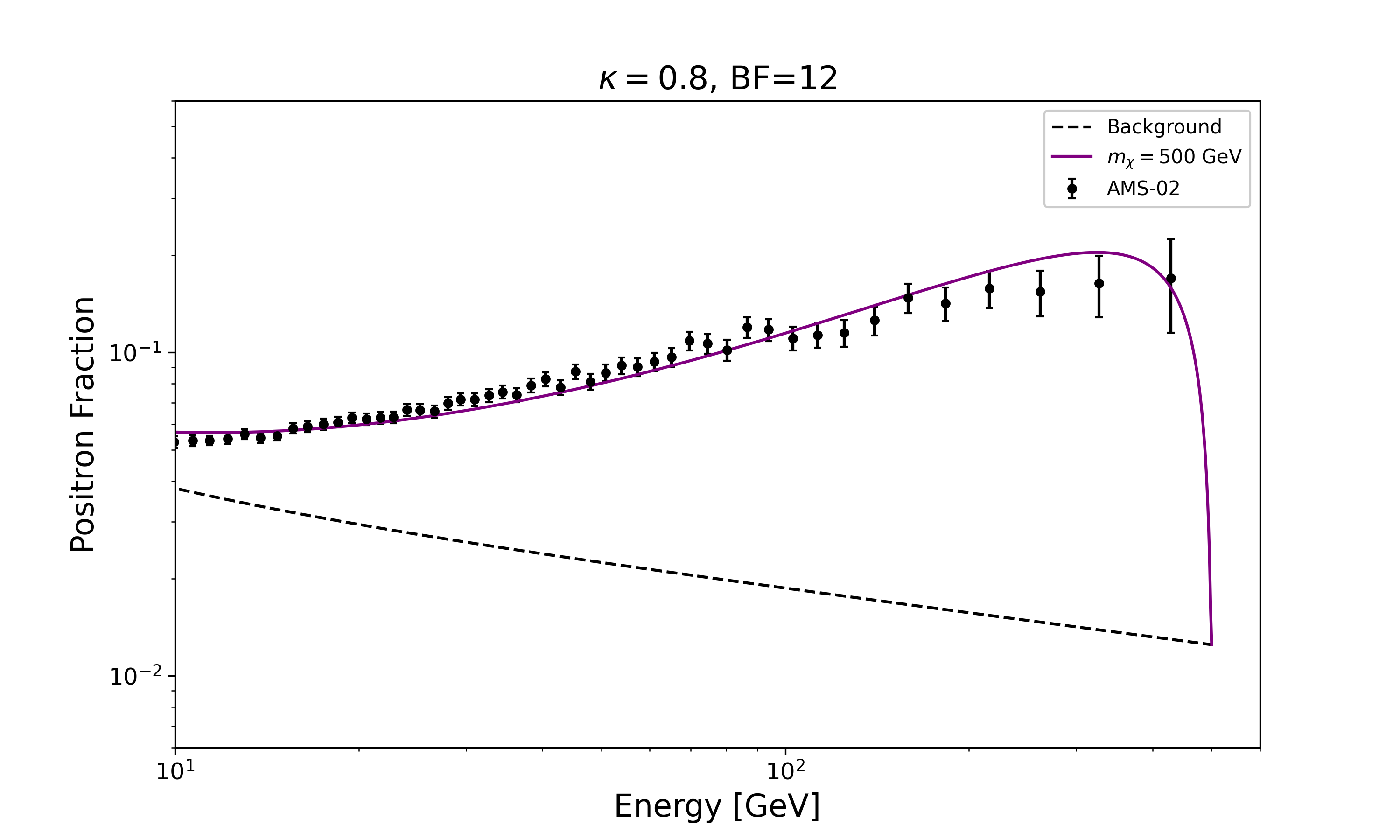}
    \end{minipage}
    \caption{Positron Flux (left) and Positron Fraction (right) as a function of energy for $m_\chi = 500$ GeV, with normalization factor $\kappa=0.8$ and boost factor $BF=12$.}
    \label{ID2}
\end{figure}

Figure~\ref{ID}-(b), shows the positron fraction $\frac{e^+}{(e^+ + e^-)}$ for the chosen values of mass. The background is plotted as a dashed line, and the positron excesses of AMS-02 is shown with uncertainty bars. For a DM mass of 100 GeV, we notice that the signal line can explain the AMS-02 data up to $E<80$ GeV, while the remaining values of mass falls below the observed data. However, it is important to note that the presented cross-sections in Table~\ref{tabid} correspond to a specific mass spectrum where the mass of the charged Higgs can reduce the computed cross-sections due to being within $\sim 100$ GeV from that of the DM particle. It is possible to change the mass spectrum through the input parameters to reduce such contributions (by pushing $m_{H^\pm}$ farther from $m_\chi$), and hence leading to an enhanced signal for the cases (since $\langle \sigma v \rangle$ will increase). Moreover, in the current setup, we have set the normalization factor $\kappa$ \cite{Baltz:1998xv} and the boost factor $BF$ to 1. It is known that such factors might influence the fluxes significantly. In particular, Figure~\ref{ID2} shows the case of $m_\chi = 500$ GeV, with $\kappa=0.8$ and $BF=12$, which can explain the AMS-02 data.

\section{Conclusion} \label{discon}

In this paper, we have extended the Inert Doublet Model (IDM) with a \( U(1)_{B-L} \) gauge symmetry to address both neutrino mass generation and dark matter (DM) abundance in the intermediate-mass range. This symmetry introduces right-handed neutrinos, providing a natural origin for neutrino masses while also offering a framework for non-thermal DM production.

We demonstrated that, within this extension, the CP-even component of the inert doublet can act as a viable DM candidate. Although the thermal relic abundance of this candidate is insufficient to match observational data, the non-thermal production mechanism we proposed, involving the decay of a heavy scalar associated with \( U(1)_{B-L} \), effectively achieves the observed DM abundance at low reheating temperatures. This pathway provides a robust alternative to thermal production, accommodating the intermediate-mass range that thermal models fail to explain.

We also explored the implications of this model for direct and indirect DM detection. Our findings indicate that the extended IDM can satisfy current experimental constraints, offering realistic prospects for future detection efforts. 

Overall, the \( U(1)_{B-L} \)-extended IDM not only addresses the shortcomings of thermal DM production but also provides a coherent framework for integrating neutrino masses, demonstrating its potential as a unified solution to key challenges in particle physics and cosmology. Future work could further refine these predictions and explore additional phenomenological implications of this model.

\section*{Acknowledgement}

The work of MB is supported by King Saud University. The work of SK is partially supported by Science, Technology \& Innovation Funding Authority (STDF) under grant number 48173.


\end{document}